# Integrated distance sampling models for simple point counts

**Short title:** Integrated distance sampling

Marc Kéry[1], J. Andrew Royle[2], Tyler Hallman[1,5], W. Douglas Robinson[3],

Nicolas Strebel[1], Kenneth F. Kellner[4]

[1] Swiss Ornithological Institute, Seerose 1, 6204 Sempach, Switzerland (marc.kery@vogelwarte.ch)

[2] U.S. Geological Survey Eastern Ecological Science Center, Laurel, 20708 Maryland

[3] Oak Creek Lab of Biology, Department of Fisheries, Wildlife, and Conservation Science, Oregon State University, Corvallis, Oregon USA;

[4] Department of Fisheries and Wildlife, Michigan State University, East Lansing, Michigan USA;

[5] Department of Biology and Chemistry, Queens University of Charlotte, Charlotte, North Carolina USA

**Open Research**

The dev version of R package `unmarked` which contains the IDS model fitting function `IDS()` is available on GitHub and can be installed from R by submitting the following command: `remotes::install_github("kenkellner/unmarked", ref="IDS")`.

For the case study, eBird data was downloaded as described in the manuscript and in Appendix S2: Section S2. Data and code used in the case study, along with R and JAGS code for conducting Simulation studies 1–4 described in the manuscript, will be posted on Zenodo upon acceptance; they are currently on GitHub: `https://github.com/kenkellner/IDS`.

# Abstract

Point counts (PCs) are widely used in biodiversity surveys, but despite numerous advantages, simple PCs suffer from several problems: detectability, and therefore abundance, is unknown; systematic spatiotemporal variation in detectability produces biased inferences, and unknown survey area prevents formal density estimation and scaling-up to the landscape level. We introduce integrated distance sampling (IDS) models that combine distance sampling (DS) with simple PC or detection/nondetection (DND) data and capitalize on the strengths and mitigate the weaknesses of each data type. Key to IDS models is the view of simple PC and DND data as aggregations of latent DS surveys that observe the same underlying density process. This enables estimation of separate detection functions, along with distinct covariate effects, for all data types. Additional information from repeat or time-removal surveys, or variable survey duration, enables separate estimation of the availability and perceptibility components of detectability. IDS models reconcile spatial and temporal mismatches among data sets and solve the above-mentioned problems of simple PC and DND data. To fit IDS models, we provide JAGS code and the new `IDS()` function in the R package `unmarked`. Extant citizen-science data generally lack adjustments for detection biases, but IDS models address this shortcoming, thus greatly extending the utility and reach of these data. In addition, they enable formal density estimation in hybrid designs, which efficiently combine distance sampling with distance-free, point-based PC or DND surveys. We believe that IDS models have considerable scope in ecology, management, and monitoring.

# Introduction

Point count methods are among the most widely used and longest-standing protocols in wildlife surveys worldwide (Rosenstock et al. 2002; Darras et al. 2021). Simple point counts (PC) are short surveys in which observers count all individuals of some set of species (single species to entire communities) detected without distance constraints (i.e., unlimited distance), or within a predefined distance from the observer. Point count methods are logistically uncomplicated and thus have been adopted in the design of countless surveys worldwide, e.g., the North American Breeding Bird Survey/BBS (Sauer et al. 2017), and many European national BBS or bird atlas schemes (Balmer et al. 2013). In addition, a wide range of what in essence are point count methods, although varying from highly standardized to essentially design-free, is at the core of rapidly growing citizen-science projects such as eBird (Sullivan et al. 2009).

Despite the prevalence of simple point counts, their simplicity is not without drawbacks; e.g., it is not possible to estimate true abundance or occupancy if visits to points are unreplicated (but see Lele et al. 2012 and Sólymos et al. 2012). In addition, PC data are non-spatial in the sense that the area from which the detected animals are drawn is unknown. This prevents spatial extrapolation for rigorous estimation of regional population sizes. Similarly, integrated analyses of data from different schemes is hampered due to commonly occurring spatial mismatches. Finally, variable survey duration is commonplace and creates temporal mismatches in the data (Pacifici et al. 2019). This further complicates joint analyses from multiple survey schemes that use point count methods.

In planned surveys, additional information is often collected during PC surveys that permits estimation of detection probability (Nichols et al. 2009). Such extra information includes replicated counts (Royle 2004), double-observer surveys (Nichols et al. 2000), removal counts (Wyatt 2002; Dorazio et al. 2005), distance information (Marquez et al. 2007; Buckland et al. 2015), or locational information from recognizable individuals which enables spatial capture-recapture (SCR) models to be fit (Borchers and Efford 2008; Royle et al. 2014). These survey protocols permit estimation of abundance, and thus assessment of status and trends free from any bias produced by imperfect detection and by unmodelled temporal or spatial patterns in detectability (Kéry & Royle 2016, 2021). However, apart from SCR and distance sampling (DS), these methods only produce detectability-corrected indices of abundance that are not area-explicit.

Even if DS or SCR data are available, it is not clear at present how they should be used alongside, or combined with, data from simple PCs that often are available from national BBS or bird atlas schemes. For instance, one major challenge in the joint modeling of such data types is how to deal with spatial or temporal mismatches (Pacifici et al. 2019), i.e., when effective sampling areas are unknown and vary, and when survey durations differ (Solymos et al. 2013). Thus, it would be desirable to have formal methods combining these different data types that build on their complementary strengths (e.g., detectability estimation, trained observers *vs.* large sample size, geographic breadth, etc.).

Here, we introduce integrated distance sampling (IDS) models that permit the combination of data from DS, simple PC, and detection/nondetection (DND) surveys. Our integrated model is based on an underlying hierarchical distance sampling model (Royle et al. 2004; Kéry & Royle 2016: chapter 8) for all three data types. We conceptualize data from all of them as the outcome from a (possibly latent) distance sampling protocol. This enables us to estimate separate detection functions for each data set, which automatically reconciles any spatial mismatch among the data types and surveys. In addition, temporal mismatches such as variable survey duration can be addressed by including in the model an availability process, which is informed by extra data such as variable survey duration, or by multi-observer, replicate or time-removal surveys (Borchers et al. 1998, Solymos et al. 2013; Amundson et al. 2014). These extra data allow for separation of the availability and perceptibility components of detection probability (Marsh & Sinclair 1989; Nichols et al. 2009, Hostetter et al. 2019, Péron & Garel 2019).

Key to our IDS models is the view of PC and DND data as aggregations, or summaries, of latent DS survey data that observe the same density, and possibly also availability, processes as the regular DS surveys. Hence, we view PC and DND data types as DS counts where distance information is unavailable. As we will show, the assumption of a shared density and availability process permits estimation of separate detection functions, along with different parameters linking these functions to covariates, for all three data types. Estimation of separate detection functions, when needed, can accommodate any systematic differences between survey schemes. Combining PC and DND data with the more information-rich DS data not only enables estimation of detection probability, but also makes the resulting abundance estimates area-explicit: effective survey areas for PC and DND surveys become estimable, and population density is estimated with improved precision. Thus, IDS models can reconcile all discrepancies, including spatial and temporal mismatches, among these widespread data types.

In this article, we begin by formally describing IDS models. Then, we use simulation to first demonstrate identifiability of our model when separate detection functions are estimated for each data type, including separate parameters for detection function covariates. Next, we explore the effects of adding variable amounts of the more information-rich but more "expensive" DS surveys to a larger sample of the less information-rich but "cheaper" PC data. Following that, we demonstrate the identifiability of a model that includes availability in addition to perceptibility, provided that survey duration is variable; variable survey duration is one of the types of extra information which enable availability in addition to perceptibility to be estimated (e.g., Solymos et al. 2013). Finally, we showcase IDS models with real data, using the Oregon 2020 Bird Survey Project (Robinson et al. 2020) as a case study. As part of the case study, we demonstrate the ability of IDS models to allow for different levels of heterogeneity in the detection functions estimated for different portions of the data (Oedekoven et al. 2015). Such accommodation of survey-specific features of the observation process will be particularly important when reconciling data from heterogeneous survey protocols in a single integrated model.

We have implemented a range of IDS models in the new fitting function `IDS()` in the R package `unmarked` (Fiske & Chandler 2011), to permit model fitting by maximum likelihood, and we provide BUGS code for Bayesian inference using JAGS (Plummer 2003). We believe that IDS models have a large scope of application for exploiting PC and DND data in a more rigorous and synthetic manner, and to obtain less biased and larger-scale inferences about abundance and density, particularly for large amounts of citizen-science data.

**Integrated Distance Sampling (IDS) models**

We develop joint likelihoods, i.e., integrated models (Besbeas et al. 2002; Miller et al. 2019; Kéry & Royle 2021: chapter 10; Schaub and Kéry 2022), for the following data types, which we assume to observe the same density and availability processes. We note that $i$ indexes different sites across data types. Also, our current models assume closure and the absence of any temporal replicates at a site, though both assumptions may be relaxed in future developments.

(1) <u>Distance-sampling (DS) data</u> $y_{i,j}^{ds}$, possibly with truncation distance $b_i^{ds}$ and survey duration $t_i^{ds}$, where $j$ indexes $J$ distance classes, and where $y_{i,\cdot}^{ds} = \sum_{j=1}^{J} y_{i,j}^{ds}$ denotes the total count per site.

(2) <u>Simple point counts (PC)</u> $y_i^{pc}$ with duration $t_i^{pc}$, with or without a truncation distance $b_i^{pc}$, as produced by many national BBS or bird Atlas schemes.

(3) <u>Detection/nondetection (DND) data</u> $y_i^{dnd}$, indicating the observed presence or absence of a species during a point-location survey of duration $t_i^{dnd}$ out to an optional truncation distance $b_i^{dnd}$, as they are similarly produced frequently by biological surveys.

For inference about density, first, for the DS data we adopt a hierarchical distance sampling model (Royle et al. 2004) represented by $N_i \sim Poisson(A_i \lambda_i)$ and $y_{i,\cdot} \sim Binomial(N_i, \theta_i p_i^{ds})$. $N_i$ and $y_{i,\cdot}$ are, respectively, the latent abundance and observed total count at site $i$, with survey area $A_i$ and density $\lambda_i$, while availability ($\theta$) and perceptibility ($p^{ds}$) are the two components of detection probability (Marsh and Sinclair 1989; Nichols et al. 2009). Perceptibility will primarily be a function of distance and is estimated from distance data by integrating out to distance $b_i$ a suitable detection function such as a half-normal with scale parameter $\sigma$. In turn, that truncation distance $b_i$ defines the survey area, which for a point count survey (assuming perfect detection) is $A_i = \pi b_i^2$.

This is a key advantage of DS methods, which associates $N_i$ with a well-defined area. It makes abundance estimates in a DS protocol area-explicit, in contrast to most abundance estimation protocols other than SCR (Borchers & Efford 2008; Royle et al. 2014). For songbirds, the availability probability $\theta$ will be mainly a function of singing rates (Solymos et al. 2013), which cannot be estimated from distance data alone. Hence, conventional distance sampling (CDS) requires either the assumption of perfect detection at distance 0 or else acceptance that inferences will be restricted to the available part of a population only (Buckland et al. 2015). However, availability becomes estimable in a DS model if certain extra information is collected, e.g., from multiple observers (Borchers et al. 1998), replicated surveys (Chandler et al. 2011), time-removal (Farnsworth et al. 2002), or alternatively from variable survey duration in an IDS model, as we will show.

Second, for the PC data we adopt a variant of the binomial *N*-mixture model (Royle 2004). Simple PCs are neither area-explicit, nor can detection probability be estimated without temporal replication (though see Lele et al. 2012 and Sólymos et al. 2012). This precludes estimation of survey area $A^{pc}$, availability $\theta^{pc}$, and perceptibility $p^{pc}$. However, we will show that use of PC data alongside regular DS data in an IDS model renders estimable both $A^{pc}$ and $p^{pc}$, again via the estimation of the parameters of a suitable detection function. Conceptualizing simple PC data as the outcome from latent DS surveys lets us estimate separate detection functions, along with distinct effects of covariates, for both DS and PC data when they are used as part of an IDS model. Integration of these detection functions over unlimited distance or out to some chosen truncation distance yields the average detection probability $\bar{p}_i^{pc}$ for a PC survey at site *i* and enables estimation of survey area $A^{pc}$. This lets PC surveys contribute information towards estimation of density $\lambda$. This model represents a complete, model-based reconciliation of the spatial mismatch between DS and PC data. In addition, variable survey duration $t^{pc}$ or other extra information mentioned above may render estimable availability $\theta$ and thus additionally reconcile temporal mismatches among surveys as well.

Third, for DND data we adopt a variant of the Royle-Nichols (2003) model. Here, the observed DND data are assumed to follow a Bernoulli distribution with a success probability that depends on both local abundance and parameters of the detection process: $y_i \sim Bernoulli(1 - (1 - \theta_i^{dnd} \bar{p}_i^{dnd})^{N_i})$, where $y_i$ denotes a binary DND datum at site $i$ and the other quantities are analogous to what was defined above. As for the PC data, single-visit DND data without any extra information won't normally permit parameter estimation in this model, but using DND data alongside DS data as part of an IDS model will render identifiable survey area $A_i^{dnd}$ and detection probability $\bar{p}_i^{dnd}$. Moreover, variable survey duration or similar extra-information will additionally permit estimation of availability $\theta$ while accounting, and reconciling surveys, for variable survey duration, i.e., for temporal mismatch of surveys.

Our current IDS models assume population closure, that DS data are always available, and that all data types observe the same abundance and availability processes. Hence, abundance and, if modelled explicitly, availability parameters are shared in a joint likelihood, while detection parameters can be either shared or made specific to each data type. We will show that this enables IDS models to obtain separate intercept and slope estimates in the detection function, and therefore of survey area $A$, density $\lambda$ and detection probability $p$, from unreplicated, simple PC or DND data, when these are used in an IDS model. If PC or DND data are the result of surveys with variable duration, a parametric model for the availability process may also be built. For songbirds, Solymos et al. (2013) express availability as a function of singing (or more generally, activity) rate $\phi$ and of survey duration $t$ as $\theta_i = 1 - \exp(-t_i \phi_i)$. We will show how we can also estimate availability in an IDS model, provided that survey duration is sufficiently variable and sample size is sufficiently large.

To summarize our IDS models, for regular DS data we specify likelihood $L^{ds}$ (Royle et al. 2004), for PC data $L^{pc}$ (Royle 2004), and for DND data $L^{dnd}$ (Royle & Nichols 2003). Importantly, for both PC and DND data, we assume a latent DS observation process protocol and estimate $p$ by integration of a detection function with parameters that become estimable in an IDS model. Under independence assumptions, i.e., when at most a negligible portion of sites appears in more than one data set, we define the following joint likelihoods for three variants of an IDS model: $L^{IDS1} = L^{ds} \times L^{pc}$ (which we call model IDS1) and $L^{IDS2} = L^{ds} \times L^{dnd}$ (model IDS2) for the combinations of DS with PC or DND data, and $L^{IDS3} = L^{ds} \times L^{pc} \times L^{dnd}$ (model IDS3) for the full three-way combination of data. These likelihoods can be maximized numerically to obtain MLEs, or we can place priors on their parameters and use MCMC methods to obtain Bayesian posterior inferences. See Appendix S1 for a conceptual outline of IDS models and of how they conceptualize PC and DND data as the outcome of a "latent" distance sampling observation process.

## Tests and demonstrations of integrated distance sampling models with simulated and real data

*Simulation 1: Identifiability of separate observation process parameters in IDS1 and IDS2*

To demonstrate identifiability of the new models, we analyzed simulated data sets and estimated parameters for separate detection functions in an IDS model with either DS + PC data or DS + DND data, i.e., in the IDS1 and IDS2 case. We used function `simHDS()` in the R package `AHMbook` to simulate two data sets with DS data from 250 sites, and PC or DND data from another 1000 sites. To obtain PC data, we first generated DS data, and then discarded all distance information, just retaining one count per site each, and to produce DND data we additionally quantized the resulting counts. Mean density was kept constant at 1, following our assumption of a shared density process. The scale parameter $\sigma$ in the half-normal detection function was set at 100 m for the DS data and was varied randomly between 10 and 130 m for the PC and DND data sets. Thus, the key criterion for identifiability of the new models was how well estimates of $\sigma$ matched their true values in the data simulation. In the submodel for the DS data sets, we chose a truncation distance of 200 m. In this simulation we aimed to establish the identifiability of the new models in their simplest form only. That is, we implicitly assumed availability to be 1 and did not use any covariates in either density or detection. We used JAGS (Plummer 2003) to fit IDS1 or IDS2 to 1000 data sets each.

In Simulation 1B (Appendix S2: Section S1) we continued our investigation of parameter identifiability and of estimator performance in model IDS1. We varied all of the following four settings independently according to a response-surface design: average density, detection function scale for both DS and PC data, and DS truncation distance. We again used JAGS to fit all models.

*Simulation 2: Identifiability with distinct covariate effects in the observation model*

We conducted two sets of simulations to answer the following related questions: (i) Does the IDS model allow DS and PC detection to have different covariate relationships? (ii) Are relationships still identifiable if the same covariates are related to both detection and density? We answered these questions by simulating data sets with DS and PC data from 200 and 1000 sites, respectively. Density was governed by an intercept of 1 on the natural scale and an effect of 1 of one covariate ("habitat"). The half-normal detection function $\sigma$ had an intercept of 100 and 150 m on the natural scale for DS and PC data, respectively. In a first analysis we used `simHDS()` to simulate 1000 data sets with these specifications, and where the half-normal detection function $\sigma$, on the log-scale, was affected by another covariate "wind" by independently drawing two random numbers from a Uniform(-0.5, 0.5) distribution. In the second analysis, we used a modified version of function `simHDS()` to simulate another 1000 data sets with the same specifications as above, except that now we generated log-scale effects of the *same* covariate as for density (i.e., "habitat") by independently drawing two U(-0.5, 0.5) random numbers for the DS and PC data sets as their coefficients. We used the new `IDS()` function in R package `unmarked` to fit the data-generating model. We discarded numerical failures, which we conservatively identified by standard errors that were either NA or >5, or by MLEs that were >10 times their true values.

*Simulation 3: How many DS sites are required to obtain adequate estimates of density?*

We simulated 1000 data sets with PC data from 200 sites, to which we added DS data from 1–100 sites in six mixing ratios. Density was governed by an intercept of 1 on the natural scale, with one habitat covariate with coefficient 1. The detection function $\sigma$ was 70 m in the PC and 100 m in the DS data, and we chose a truncation distance of 200 m in the latter. We generated a total of 6000 data sets (1000 for each level of the mixing ratio factor) and fit the IDS1 model using function `IDS()`, discarding numerical failures as in Simulation 2.

*Simulation 4: How well can availability be estimated in an IDS model?*

We simulated 1000 data sets that resembled our case study below: each had DS data from 3000 sites, and PC data from either 1000, 3000, or 6000 sites. DS survey duration was kept constant at 5 min, but was varied between 3 and 30 min in PC surveys, with a strong right skew, as found in the case study data. Density was goverened by an intercept of 1 on the natural scale and with a habitat covariate with coefficient 1, detection function $\sigma$ was 70 m in the PC and 100 m in the DS data, with a truncation distance of 200 m in the latter. Singing rates varied between 0.1 and 2, corresponding to a probability of 0.1–0.86 to sing at least once over a 5 min interval. We fit IDS1 using the `IDS()` function and discarded numerical failures as in Simulation 2.

*Case study: American Robins in the Oregon 2020 Bird Survey Project*

We used the IDS3 model to estimate population density of American Robin (*Turdus migratorius*) in Benton and Polk counties, Oregon. The 3680 km$^2$ area in Western Oregon is bounded on the east by the Willamette River and its floodplain, while the western portions include the Coast Range mountains. Silviculture of coniferous forest is the dominant land use in the mountains. Nearly every square kilometer contains a narrow, lightly travelled road for timber harvest, which allowed access for bird surveys. The eastern floodplain sections contain a mix of agricultural uses, mostly festucoid grass seed fields and orchards, and suburban development.

DS surveys were conducted every 0.8 km along accessible roads throughout the study area, and every 200-m in an off-road grid placed over the William L. Finley National Wildlife Refuge, producing a total of 2,912 sites sampled and 2020 American Robins detected (Robinson et al. 2020). DS surveys were conducted during the breeding season (April 30–July 11) from 2011 to 2013 by WDR. Each survey followed the Oregon 2020 protocol (Robinson et al. 2020), which used 5-minute stationary counts initiated between 30 min before sunrise and noon on days with no or little rain. All birds detected by sight or sound were recorded with an estimated distance from the observer (verified with a range finder when possible) following standard distance sampling protocols (Buckland et al. 2015).

We combined DS surveys with opportunistically-gathered citizen-science PC data from the eBird database (Sullivan et al. 2009), using checklists from 2011-2017 in Benton and Polk counties. After stringent filtering (see Appendix S2: Section S2) and geographic sampling, 1060 PCs with 819 detections of American Robins were included. We filtered data to include only complete checklists using stationary protocols and personal locations, conducted during the breeding season. We further filtered data to include only checklists with durations between 3-30 minutes that were conducted between sunrise and seven hours after sunrise. Finally, we applied geographic sampling to reduce the effects of popularly birded sites by overlaying a 200m grid over the study area and randomly selecting only a single checklist from each grid cell. To illustrate the combination of DS data with both simple PC and DND data, we randomly subset half of the PC data and reduced it to DND. This resulted in 521 simple PCs with a total of 394 individuals counted, plus 538 DNDs, among them 228 with at least one detection of the species. See Appendix S2: Section S2 for eBird query details.

For DS data, we selected a truncation distance $b^{ds}$ of 300 meters. We binned the distance data into 50 m distance classes. For the analysis of PC and DND data, an upper distance limit $b^{pc}$ and $b^{dnd}$ of 500 meters was adopted, assuming that observers do not detect individuals further away than 0.5 km (the 0.99 quantile in the Oregon 2020 database (Robinson et al. 2020) was 0.4 km). For all data types, we assumed identical parameters for annual density and availability. We modelled density $\lambda$ with a random intercept for year, and with quadratic terms for elevation and percentage of canopy cover in a 315 m radius around the observer location. This radius was selected as it was previously found to be the most predictive of abundance for this species of the radii considered (Hallman and Robinson 2020). For availability, we adopted the model of Solymos et al. (2013) linking availability probability with activity rate $\phi$, and used quadratic terms for day of the year and minutes since dawn on the log activity rate.

The observation process in the designed DS surveys in the Oregon 2020 project may differ from point count surveys recorded in eBird, even after stringent filtering. Therefore, we allowed for different detection functions for the DS and the PC/DND portions in our analysis by fitting separate intercepts in the half-normal detection scale $\sigma$. Moreover, to accommodate possibly different levels of detection heterogeneity among sites, we specified site-specific random effects in $\sigma$ with a potentially different variance for the DS and PC/DND portions of the data (Oedekoven et al. 2015). In addition, we modeled $\sigma$ using the percentage of urban area and percentage of canopy cover, both in a 165 m radius around the observer location; these slope parameters were shared between DS and PC/DND data. We used a smaller radius for canopy cover in the detection function as the distance that an observer can detect is impacted more heavily by nearby environmental conditions.

Elevation was obtained from the Oregon Spatial Data Library (Oregon Spatial Data Library 2017), urban land cover was obtained from the U.S. Geological Survey's National Gap Analysis Project (USGS 2011), and canopy cover was obtained from Landscape Ecology, Modeling, Mapping and Analysis's gradient nearest neighbor structure maps (LEMMA 2014).

We processed data in R (R Core Team 2019) and fitted the model in JAGS, using the R package `jagsUI` (Kellner 2016). For all parameters, we chose vague priors (see BUGS model on ZENODO for specifics). We assessed the goodness of fit for our model for each data portion separately using posterior predictive checks (Conn et al. 2018) with a Freeman-Tukey discrepancy measure between the observed and the expected counts for the DS and PC data (Kéry & Royle 2016). For the binary DND we used the observed number of sites with detections as a test statistic and checked whether the actual value for the DND data was within a 95% credible interval (CRI) of this statistic in data sets replicated under the model. This approach suggested an adequate fit of the model overall: Bayesian p-values for the DS part of the model revealed slight underdispersion of the data, while the PC and DND parts of the data indicated a fitting model (Appendix S2: Table S6). We obtained posterior predictive distributions of abundance to project density onto a raster map with elevation and canopy cover for each of the 3874 1-km$^2$ grid cells in Benton and Polk counties to obtain an abundance-based species distribution map of the Robin in our study area. We also fit a simpler variant of the model using the `IDS()` function in `unmarked` (Fiske & Chandler 2011) to illustrate both the Bayesian and the maximum likelihood approach. Code and data to replicate the case study can be found on Zenodo.

## Results

### *Simulation 1: Identifiability of separate observation process parameters in IDS1 and IDS2*

In an IDS model, separate detection functions can clearly be estimated under both IDS1 (combining DS and PC data) and IDS2 (combining DS and DND data); see Fig. 1 and Appendix S2: Table S1. There were no signs of bias in either model: % relative bias was <<1% for all sigma's and <2% for the abundance estimates at sites with *N*>0. Credible interval (CRI) coverage was close to the nominal level of 95% for all estimators. Not surprisingly, precision was slightly lower in model IDS2 than in IDS1 (see middle of Fig. 1). In addition, simulation 1B confirmed the good frequentist operating characteristics of the estimators in IDS models under an even wider range of conditions (Appendix S2: Section S1, Appendix S2: Table S2).

### *Simulation 2: Identifiability with distinct covariate effects in the observation model*

In the first set of simulations, where two distinct covariates affected density and the detection function, but where the effects on the latter were distinct for the DS and PC portions of the data, we discarded 23 sets of estimates as invalid. The remaining 977 sets of estimates indicated that this model was identifiable and showed little or no signs of bias (Fig. 2, left; Appendix S2: Table S3 left). In the second set of simulations, where a single covariate independently affected density and the two detection functions, we discarded 66 invalid sets of estimates. The remainder again showed this model to be identifiable (Fig. 2, right; Appendix S2: Table S3 right).

*Simulation 3: How many DS sites are required to obtain adequate estimates of density?*

The IDS model showed excellent performance with as few as 20 DS sites (Fig. 3), with relative bias <1% for all estimators and CI coverage at or near nominal levels (Appendix S2: Table S4). However, the number of numerical failures increased greatly when decreasing numbers of DS data were added in the integrated model; from only 2 out of 1000 when 100 DS sites added, to 85 with 20 DS sites, and to 490 out of 1000 when 1 DS site was added.

*Simulation 4: How well can availability be estimated in an IDS model?*

Sampling distributions of density estimators were all concentrated around the true value. There were long right tails, but these became more symmetrical with larger sample sizes. Singing rate ($\phi$) estimators were precise up to values of about 0.8, 1.3 and 1.4, respectively, for 1000, 3000 and 6000 PC sites, but became very imprecise for greater values of the singing rate. Presumably, this was because overall availability reached an asymptote close to 1 when singing rates were very high, making precise estimation of $\phi$ hard (Fig. 4). Overall, there was a slight positive bias in both density and singing rates, but it declined from 14 and 10% with 1000 PC sites to 3000 and 2% with 6000 sites, while CI coverage was always at nominal levels (Appendix S2: Table S5). Relative bias of the detection function $\sigma$ for both data types was 0.76% or less throughout.

*Case study: American Robins in Oregon*

Over all surveys considered, mean survey date was June 7, and time since dawn ranged 17–519 min (mean 229). At mean date and time since dawn, availability within a one minute survey was estimated at 0.27 (95% CRI 0.15–0.46; Appendix S2: Table S6). Estimated availability peaked soon after dawn, decreased during the next five hours, then increased again, and tended to increase slightly throughout the season. Density was estimated to be highest on plots with a canopy cover of ~40% and to decrease with elevation. Median estimates for density varied between 2 and 47 individuals per km$^2$. Maxima were found in the foothills where open woodlands transition from the floodplain agricultural zones into the denser forests at higher elevations, while minima were found in the most intensively harvested woodlands (Fig. 5). Over the entire study area, we estimated a population size of 74,344 American Robins (95% CRI 48,039–117,944). Interestingly, on average the estimated detection function was not different between the DS and PC/DND portions of the data (see parameter 'mean.sigma' in Appendix S2: Table S6). However, there was greater heterogeneity in the detection function $\sigma$ for surveys on eBird than for regular DS surveys conducted within the Oregon 2020 project (see parameter 'sd.eps').

## Discussion

We discovered how simple point count (PC) or detection/nondetection (DND) data can be formally integrated in a model together with distance sampling (DS) data, to estimate separate parameters of an underlying latent DS observation process. This allows estimation of full detection probability parameters for all three data types. Moreover, integrating DS data makes abundance estimates from PC and DND data area-explicit. Thereby, IDS models achieve a formal spatial calibration of PC and point-indexed DND data, as well as a reconciliation of spatial mismatches between all three data types. IDS models thus solve two major problems that plague simple point count surveys producing PC or DND data: that detection probability and effective survey areas are both unknown. The main assumption of our IDS model is a shared density process: that either density is identical among all sample locations, or that differences can be explained by covariate regressions with coefficients that are identical for all data types in the integrated model. These assumptions should be reasonable when all data types are collected in the same general area, and they may also hold when data sets are from disjoint regions. However, as in perhaps all cases where different data sets are combined in a single analysis, this is a judgment call on the part of the analyst.

  We believe IDS models have a large scope of application and can facilitate use of the large amounts of currently available PC data, such as the North American BBS (Sauer et al. 2017) in more formal analyses of abundance that account for imperfect detection They may also be applied for carefully quality-controlled eBird data (Sullivan et al. 2009), as illustrated in our case study. We have shown that only a relatively small amount of regular DS data are required to supplement simple PC data when used in an IDS model. Our findings agree with related work with other types of integrated models that demonstrate the benefits of combining even small amounts of data with a higher information content with less informative, but cheaper data (Dorazio 2014, Zipkin et al. 2017, Doser et al. 2021). This suggests that the scope of inference of point count surveys may be substantially increased by adding even a relatively small number of sites where the additional distance information is collected.

In our case study we found that perceptibility was not different on average between the DS data contributed by the Oregon 2020 project and the PC data obtained on eBird: the intercepts of the detection function $\sigma$ were hardly different between these two portions of the data used in the IDS model. However, allowing for random variation of the detection function $\sigma$ between surveys (i.e., sites) and for possibly different magnitudes of that variation between the two types of data revealed greater heterogeneity among surveys in the eBird data base than among surveys in the Oregon 2020 project. Arguably, this makes intuitive sense, since it appears likely that consistency between surveys was higher in the latter than in eBird. In addition, our case study emphasizes how careful modeling of patterns in the detection function of an IDS model can help to make data from different protocols more 'comparable', or 'alike', by explicitly allowing for their differences in terms of the observation processes that produce them. This is a great strength of IDS models and of parametric statistical inference in general.

Many survey data typically have large variation in duration (Solymos et al. 2013) and thus there is also a need for temporal mismatch among datasets to be addressed. We conceive of this as an availability process (Kendall et al. 1997), where over time some activity such as singing puts individuals at an increasing risk of being detected. Hence, survey duration is naturally informative about availability. However, this part of our model presents more challenges. Population closure is required and hence surveys should probably not be very long in duration. More importantly, this part of our IDS models has the form of a single-visit occupancy or N-mixture model (Lele et al. 2012): estimability hinges upon a continuous, "private" covariate that affects detection, and in our case, availability. Such models are identifiable (Dorazio 2012), but they rely strongly on parametric assumptions and may lack robustness to violations of those assumptions (Knape & Korner-Nievergelt 2015). Our simulations and the case study both showed availability to be identifiable in an IDS model. However, our study species was chosen to be fairly common. In contrast, in rarer species, there may well be challenges when attempting to estimate that parameter. When information to estimate availability is too sparse, estimates may tend towards the boundary of 1; however, this is what any model ignoring availability assumes implicitly anyway.

Therefore, IDS models that estimate availability must be developed and applied with care. Any extra information about availability should be incorporated in the model, such as data from multiple observers (as in mark-recapture distance sampling; Borchers et al. 1998), replicated surveys (Chandler et al. 2011), time-of-detection and time-removal data (Farnsworth et al. 2005; Alldredge et al. 2007; Solymos et al. 2013, Amundson et al. 2014). Alternatively, availability parameters may be estimated from altogether different data types, such as recordings of individual singing behavior, or perhaps even taken from the literature. We note that Solymos et al. (2013) had good success with the integration of time-removal and distance sampling data, but in a simpler model that did not involve estimation of a detection function for the time-removal data.

We can envision at least four major extensions to the IDS models described in this paper. First is the accommodation of survey sites included in the dataset which were sampled using multiple protocols. This induces a dependence that must be addressed in the construction of the joint likelihood. Second, IDS models could be developed for other survey geometries, such as line transects or search-encounter designs (Royle et al. 2014, Mizel et al. 2018). Third, allowing for open populations and demographic processes (Kéry & Royle, 2021: chapters 1 and 2) will be an important extension that may open up avenues for truly large-scale demography models (see also Appendix S1). Fourth, additional data types may be incorporated in the model, such as opportunistic data conceptualized as point patterns (Farr et al. 2021), time-to-detection data (Strebel et al. 2021), aggregated counts (Schmidt et al. 2022), and data from autonomous recording units (ARUs) (Doser et al. 2021). For instance, IDS models may be beneficial for ARU data by allowing estimation of the "listening range" of these devices under widely varying conditions, while additionally exploiting the information on singing rate contributed by the ARU data.

In summary, we believe that IDS models have a substantial scope to improve analyses of widely available simple PC and DND data obtained in citizen-science schemes, as well as the increasing amount of ARU data. IDS models may serve as a keystone of the formal, model-based unification of various data types both from designed and less-designed to even design-free surveys, to great mutual benefit. We find it fascinating to see how DS and simple PC or DND data both contribute two essential pieces of information towards the full IDS model: DS data contain most information about the detection function, while the heterogeneity in survey duration commonly found in simple PC/DND data enables estimation of the availability process. This neatly illustrates the fact that the future of biodiversity monitoring arguably lies in a combination of both designed surveys and carefully chosen citizen-science schemes.

**Acknowledgements:** The Bob and Phyllis Mace Professorship to WDR supported the Oregon 2020 project, while Reto Burri commented on an earlier draft. We thank two reviewers of an earlier draft of this paper, as well as the associate editor, for extremely helpful comments. Any use of trade, product, or firm names is for descriptive purposes only and does not imply endorsement by the U.S. Government.

## References

Alldredge, M.W., Simons, T.R., Pollock, K.H. 2007. Factors affecting aural detections of songbirds. *Ecological Applications*, 17, 948–955.

Amundson, C.L., Royle, J.A., Handel, C.M. 2014. A hierarchical model combining distance sampling and time removal to estimate detection probability during avian point counts. *Auk*, 131, 476–494. a

Balmer, D.E., Gillings, S., Caffrey, B.J., Swann, B., Downie, I.S., Fuller, R.J. 2013. *Bird Atlas 2007–11: The Breeding and Wintering Birds of Britain and Ireland*. Thetford: BTO Books.

Besbeas, P., Freeman, S.N., Morgan, B.J.T., Catchpole, E.A. 2002. Integrating mark-recapture-recovery and census data to estimate animal abundance and demographic parameters. *Biometrics*, 58, 540–547.

Borchers, D.L., Efford, M.G. 2008. Spatially explicit maximum likelihood methods for capture-recapture studies. *Biometrics*, 64, 377–385.

Borchers, D.L., Zucchini, W., Fewster, R.M. 1998. Mark-recapture models for line transect surveys. *Biometrics*, 54, 1207–1220.

Buckland, S.T., Rexstad, E.A., Marques, T.A., Oedekoven, C.S. 2015. *Distance sampling: methods and applications.* Springer, Cham, Switzerland.

Chandler, R.B., Royle, J.A., King, D.I., 2011. Inference about density and temporary emigration in unmarked populations. *Ecology*, 92, 1429–1435.

Conn, P.B., Johnson, D.S., Williams, P.J., Melin, S.R., Hooten, M.B. 2018. A guide to Bayesian model checking for ecologists. *Ecological Monographs*, 88, 526–542.

Darras, K.F.A., Yusti, E., Huang, J.C.-C, Zemp, D.-C., Kartono, A.P., Wanger, T.C. 2021. Bat point counts: A novel sampling method shines light on flying bat communities. *Ecology and Evolution*, 11, 17179-17190

Dorazio, R.M., Jelks, H.L., Jordan, F. 2005. Improving removal-based estimates of abundance by sampling a population of spatially distinct subpopulations. *Biometrics*, 61, 1093–1101.

Dorazio, R.M. 2012. Predicting the geographic distribution of a species from presence-only data subject to detection errors. *Biometrics*, 68, 1303–1312.

Dorazio, R.M. 2014. Accounting for imperfect detection and survey bias in statistical analysis of presence-only data. *Global Ecology and Biogeography*, 23, 1472–1484.

Doser, J.W., Finley, A.O., Weed, A.S., Zipkin. E.F. 2021. Integrating automated acoustic vocalization data and point count surveys for estimation of bird abundance. *Methods in Ecology and Evolution*, 12, 1040–1049.

Elith, J., Kearney, M. Phillips, S. 2010. The art of modelling range-shifting species. *Methods in Ecology and Evolution*, 1, 330-342.

Farnsworth, G.L., Nichols, J.D., Sauer, J.R., Fancy, S.G., Pollock, K.H., Shriner, S.A., Simons, T.R. 2005. *Statistical Approaches to the Analysis of Point Count Data: A Little Extra Information Can Go a Long Way*. USDA Forest Service Gen. Tech. Rep. PSW-GTR-191.

Farr, M.T., D.S. Green, K.E. Holecamp, E.F. Zipkin. 2021. Integrating distance sampling and presence-only to estimate species abundance. *Ecology*, 102, e03204.

Fiske, I., Chandler, R. 2011. unmarked: an R package for fitting hierarchical models of wildlife occurrence and abundance. *Journal of Statistical Software*, 43, 1–23.

Hallman, T.A., Robinson, W.D. 2020. Comparing multi- and single-scale species distribution and abundance models built with the boosted regression tree algorithm. *Landscape Ecology*, 35, 1161–1174.

Hostetter, N.J., Gardner, B., Sillett, T.S., Pollock, K.H., Simons, T.R. 2019. An integrated model decomposing the components of detection probability and abundance in unmarked populations. *Ecosphere*, 10(3).

Kellner, K. 2016. jagsUI: A Wrapper Around 'rjags' to Streamline 'JAGS' Analyses. R package version 1.4.4. CRAN.R-project.org/package=jagsUI

Kendall, W.L., Nichols, J.D., Hines, J.E. 1997. Estimating temporary emigration using capture–recapture data with Pollock's robust design. Ecology, 78, 563–578.

Kéry, M., Royle, J.A. 2016. *AHM1—Modeling distribution, abundance and species richness using R and BUGS. Volume 1: Prelude and Static Models*. Elsevier / Academic Press.

Kéry, M., Royle, J.A. 2021. *AHM2—Modeling distribution, abundance and species richness using R and BUGS. Volume 2: Dynamic and Advanced Models*. Elsevier / Academic Press.

Knape, J., Korner-Nievergelt, F. 2015. Estimates from non-replicated population surveys rely on critical assumptions. *Methods in Ecology and Evolution*, 6, 298–306.

Lele, S.R., Moreno, M., Bayne, E. 2012. Dealing with detection error in site occupancy surveys: what can we do with a single survey? *Journal of Plant Ecology*, 5, 22–31.

LEMMA. 2014. Landscape Ecology, Modeling, Mapping, and Analysis LEMMA. GNN Structure Maps. https://lemma.forestry.oregonstate.edu/data/structure-maps. Accessed 6 Sep 2016

Marques, T.A., Thomas, L., Fancy, S.G., Buckland, S.T., 2007. Improving estimates of bird density using multiple-covariate distance sampling. The Auk, 124(4), 1229-1243.

Marsh, H., Sinclair, D.F. 1989. Correcting for visibility bias in strip transect aerial surveys of aquatic fauna. *Journal of Wildlife Management*, 53, 1017–1024.

Miller, D.A.W., Pacifici, K., Sanderlin, J.S., Reich, B.J. 2019. The recent past and promising future for data integration methods to estimate species' distributions. *Methods in Ecology and Evolution*, 10, 22–37.

Mizel, J.D., Schmidt, J.H., Lindberg, M.S. 2018. Accommodating temporary emigration in spatial distance sampling models. *Journal of Applied Ecology*, 55, 1456–1464.

Nichols, J.D., Hines, J.E., Sauer, J.R., Fallon, F.W., Fallon, J.E., Heglund, P.J. 2000. A double-observer approach for estimating detection probability and abundance from point counts. *Auk*, 117, 393–408.

Nichols, J.D., Thomas, L., Conn, P.B. 2009. Inferences about landbird abundance from count data: recent advances and future directions. pp. 201–235 in D.L. Thomson, E.G. Cooch, M.J. Conroy (eds.) *Modeling demographic processes in marked populations*. Springer.

Oedekoven, C.S., Laake, J.L., Skaug, H.L. 2015. Distance sampling with a random scale detection function. *Environmental and Ecological Statistics*, 22, 725–737.

Oregon Spatial Data Library. 2017. Oregon 10m Digital Elevation Model (DEM). http://spatialdata.oregonexplorer.info/geoportal/details;id=7a82c1be50504f56a9d49d13c7b49aa. Accessed 30 Nov 2015

Pacifici, K., B.J. Reich, D.A.W. Miller, B.S. Pease. 2019. Resolving misaligned spatial data with integrated species distribution models. *Ecology*, 100, e02709.

Péron, G., Garel, M. 2019. Analzying patterns of population dynamics using repeated population surveys with three types of detection data. *Ecological Indicators*, 106, 105546.

Plummer, M. 2003. JAGS: a program for analysis of Bayesian graphical models using Gibbs sampling. Proceedings of the 3rd International Workshop in Distributed Statistical Computing (DSC 2003) (eds K. Hornik, F. Leisch& A. Zeileis), 1–10, TU Vienna, Austria.

R Core Team. 2019. *R: A language and environment for statistical computing.* R Foundation for Statistical Computing, Vienna, Austria. https://www.R-project.org/

Robinson, W.D., T.A. Hallman, J.R. Curtis. 2020. Benchmarking the avian diversity of Oregon in an era of rapid change. *Northwestern Naturalist*, 101, 180–193.

Robinson, W.D., T.A. Hallman, R.A. Hutchinson. 2021. Benchmark bird surveys help quantify counting accuracy in a citizen-science database. *Frontiers in Ecology and Evolution*, 9, https://doi.org/10.3389/fevo.2021.568278.

Rosenstock, S.S., Anderson, D.R., Giesen, K.M., Leukering, T., Carter, M.F. 2002. Landbird counting techniques: current practices and an alternative. *Auk*, 119, 46–53.

Royle, J.A. 2004. N-mixture models for estimating population size from spatially replicated counts. *Biometrics*, 60, 108–115.

Royle, J.A., Chandler, R.B., Sollmann, R., Gardner, B., 2014. *Spatial Capture-Recapture*. Academic Press.

Royle, J.A., Dawson, D.K., Bates, S. 2004. Modeling abundance effects in distance sampling. *Ecology*, 85, 1591–1597.

Royle, J.A., Nichols, J.D. 2003. Estimating abundance from repeated presence-absence data or point counts. *Ecology*, 84, 777–790.

Sauer, J.R., Pardieck, K.L., Ziolkowski Jr, D.J., Smith, A.C., Hudson, M.A.R., Rodriguez, V., Berlanga, H., Niven, D.K., Link, W.A. 2017. The first 50 years of the North American breeding bird survey. *The Condor: Ornithological Applications*, 119, 576–593.

Schaub, M., Kéry, M. 2022. *Integrated Population Models — Theory and ecological applications with R and JAGS.* Elsevier, Academic Press.

Schmidt, J.H., Wilson, T.L., Thompson, W.L., Mangipane, B.A. 2022. Integrating distance sampling survey data with population indices to separate trends in abundance and temporary immigration. *J. Wildlife Management*, 86, e22185.

Sólymos, P., Matsuoka, S.M., Bayne, E.M., Lele, S.R., Fontaine, P., Cumming, S.G., Stralberg, D., Schmiegelow, F.K.A., Song, S.J. 2013. Calibrating indices of avian density from non-standardized survey data: making the most of a messy situation. *Methods in Ecology and Evolution*, 4, 1047–1058.

Strebel, N., Fiss, C.J., Kellner, K.F., Larkin, J.L., Kéry, M., Cohen, J. 2021. Estimating abundance based on time-to-detection data. *Methods in Ecology and Evolution*, 12, 909–920.

Sullivan, B.L., C.L. Wood, M.J. Iliff, R.E. Bonney, D. Fink, S. Kelling. 2009. eBird: A citizen-based bird observation network in the biological sciences. *Biological Conservation*, 142, 2282–2292.

USGS (U.S. Geological Survey). 2011. National Gap Analysis Project. https://gapanalysis.usgs.gov/gaplandcover/data/download/. Accessed 7 Dec 2015

Wyatt, R.J. 2002. Estimating riverine fish population size from single- and multiple-pass removal sampling using a hierarchical model. *Can. J. of Fish. and Aquatic Sciences*, 59, 695–706.

Zipkin, E.F., Rossman, S., Yackulic, C.B., Wiens, J.D., Thorson, J.T., Davis, R.J., Grant, E.H.C. 2017. Integrating count and detection–nondetection data to model population dynamics. *Ecology*, 98(6), 1640-1650.

**Figure captions**

Fig. 1: Simulation-based validation of two integrated distance sampling (IDS) models (Simulation 1). Left: Model IDS1 (= distance sampling (DS) + simple point count (PC) data), right: Model IDS2 (=DS + detection/nondetection (DND) data); see main text for details. Top: estimation error in detection function sigma ($\sigma$) in the DS data (n = 250 sites); middle: estimated (with 95% CRIs) *vs.* true value of $\sigma$ in the PC and the DND data sets (n = 1000 sites); bottom: estimation error in the latent site-level abundances (*N*) in the PC and the DND data (mean/sd of simulated true abundance: 79/9). Red denotes truth or absence of estimation error, dashed blue shows mean of estimates. Sample size in both simulations is 1000 data sets. See also Appendix S2: Figure S1 and Tables S1-S2.

Fig. 2: Another simulation-based validation of IDS1 combining DS and PC data (Simulation 2). Left: Sampling distributions of intercept and slope estimates for detection function parameters with independent effects in the distance sampling (top) and the point count (bottom) parts of the data (Simulation 2a). Right: Intercept and slope estimates for detection function parameters with independent effects in the distance sampling (top) and the point count (bottom) parts of the data, when the same covariate has also an effect on density (Simulation 2b). Red denotes truth, dashed blue shows mean of estimates. Sample size in both simulations is 1000 data sets. See also Appendix S2: Table S3.

Fig. 3: Sampling distributions of estimators of density (intercept and slope of a continuous covariate, shared between distance sampling (DS) and simple point count (PC) data), and of detection function sigma ($\sigma$) for the DS and the PC parts of the data (Simulation 3). Throughout, sample size for the simple PCs is 200 and true values are indicated with dashed red lines. Each individual boxplot summarizes between 515 and 998 data sets that resulted in valid estimates, see also Appendix S2: Table S4.

Fig. 4: Sampling distributions of estimators of density (lambda, $\lambda$) and of activity/singing rate (phi, $\phi$) in an IDS model with availability ($\phi$) fit to data from 3000 DS sites, plus 1000, 3000, or 6000 PC sites added (Simulation 4, n = 930, 866, and 997 valid analyses). Red denotes truth or absence of estimation error, dashed blue shows mean of valid estimates. See also Appendix S2: Table S5.

Fig. 5: Estimated density of American Robin (individuals per 1 km$^2$) in Benton and Polk counties, state of Oregon, USA, based on breeding season observation data from 2011 to 2017.

Figure 1

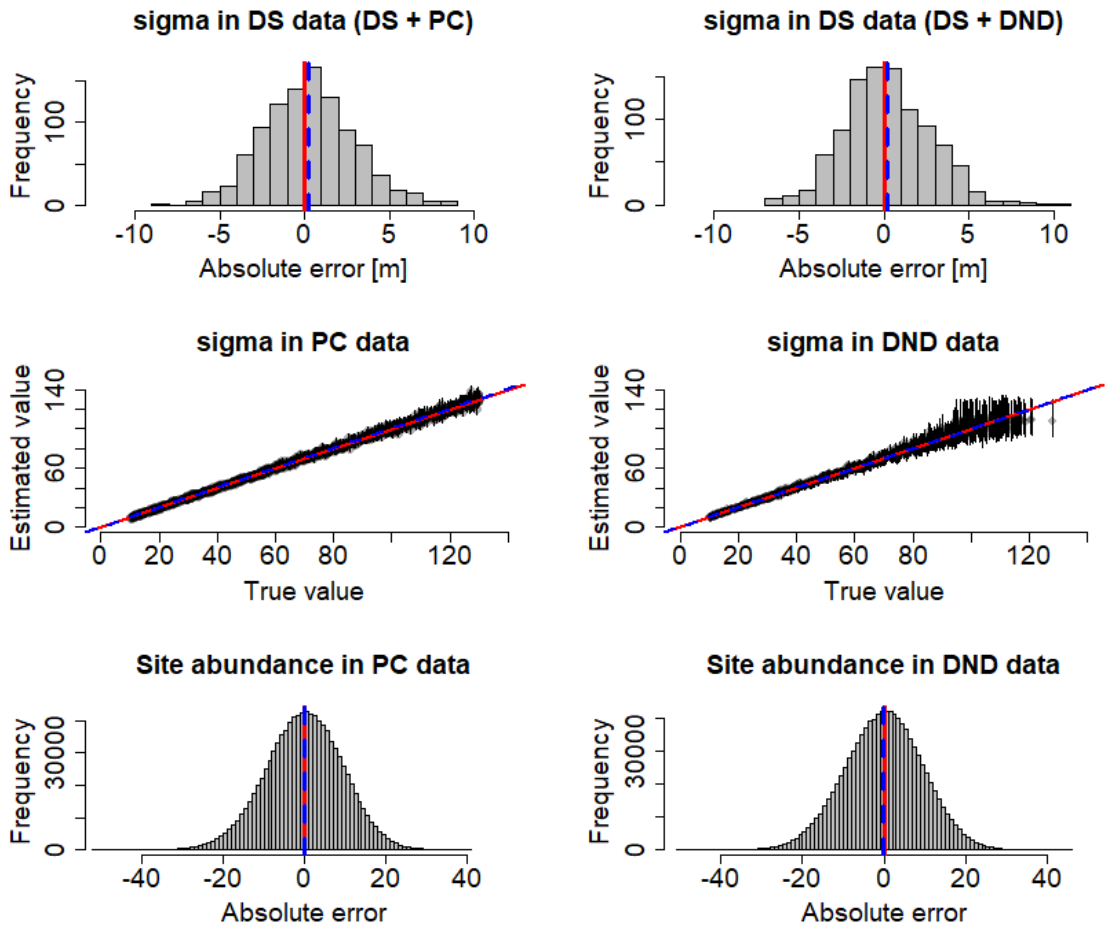

Figure 2

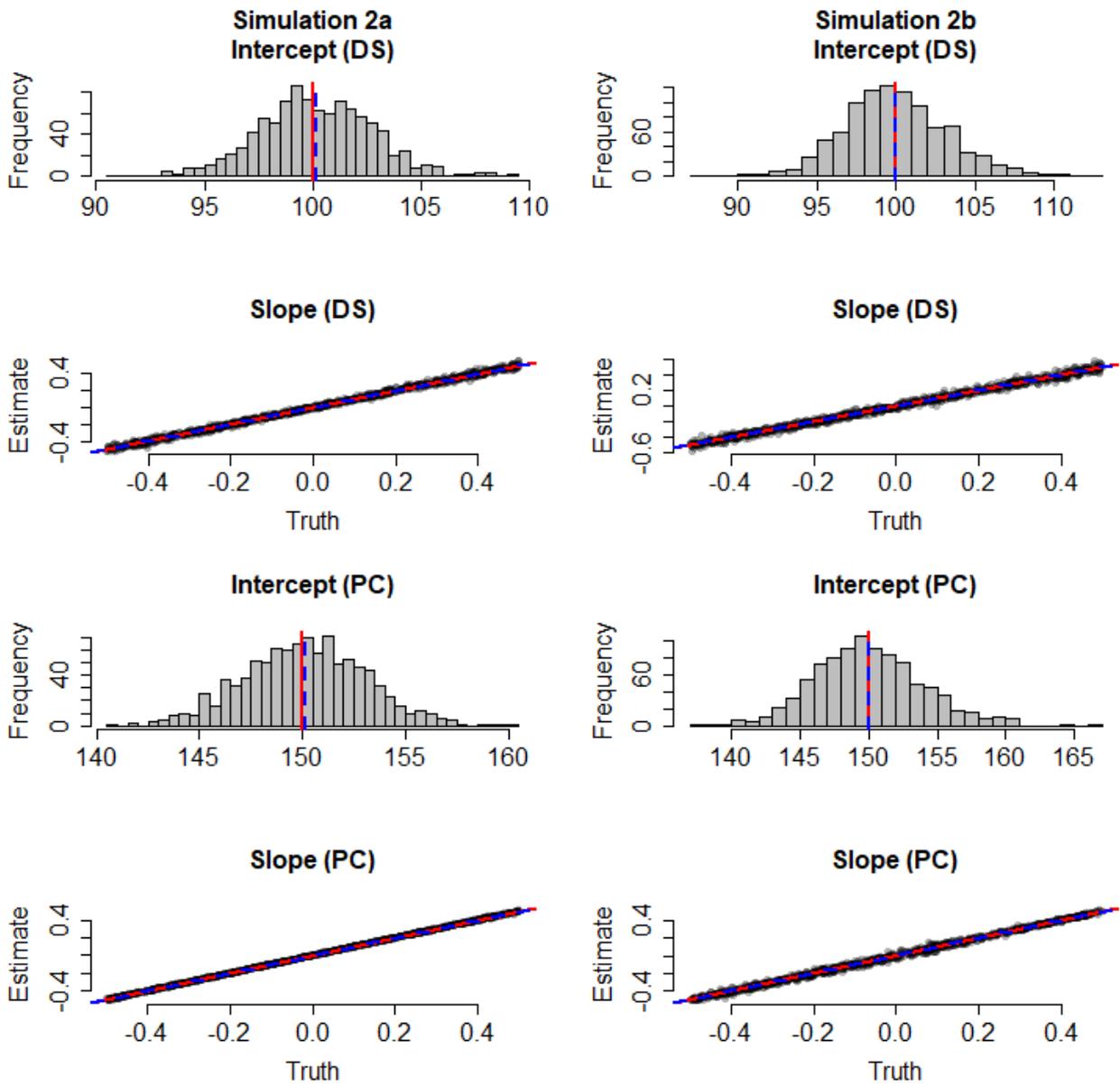

Figure 3

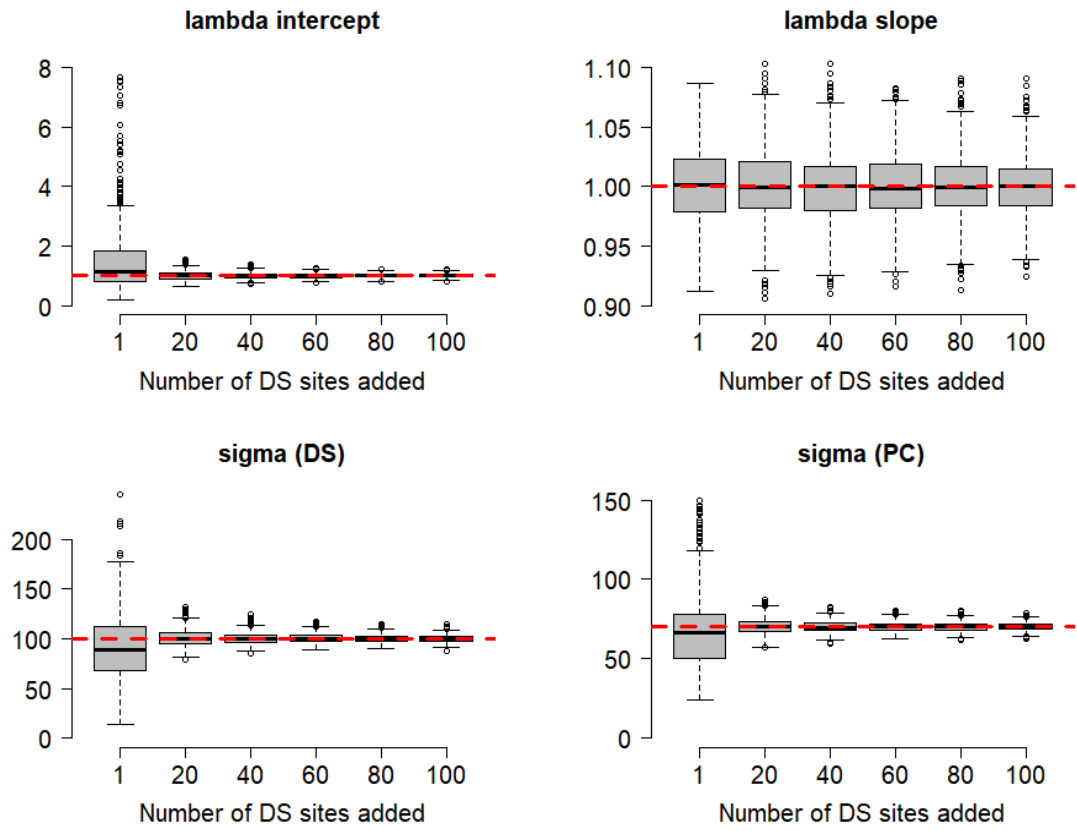

Figure 4

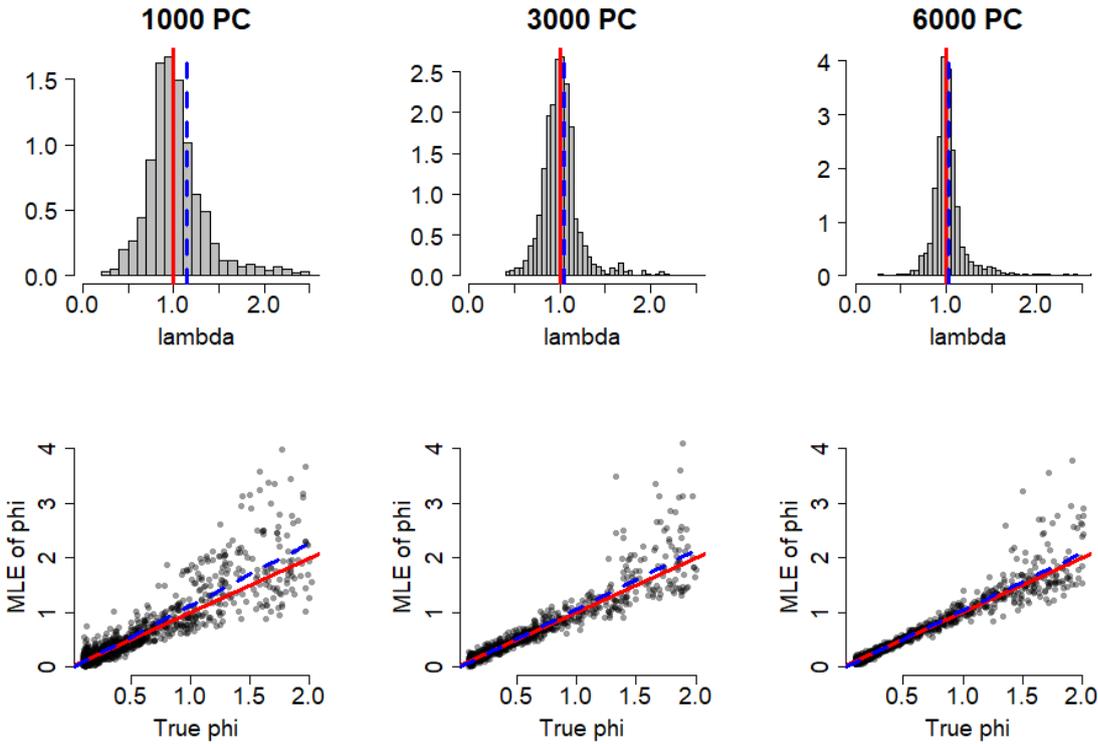

Figure 5

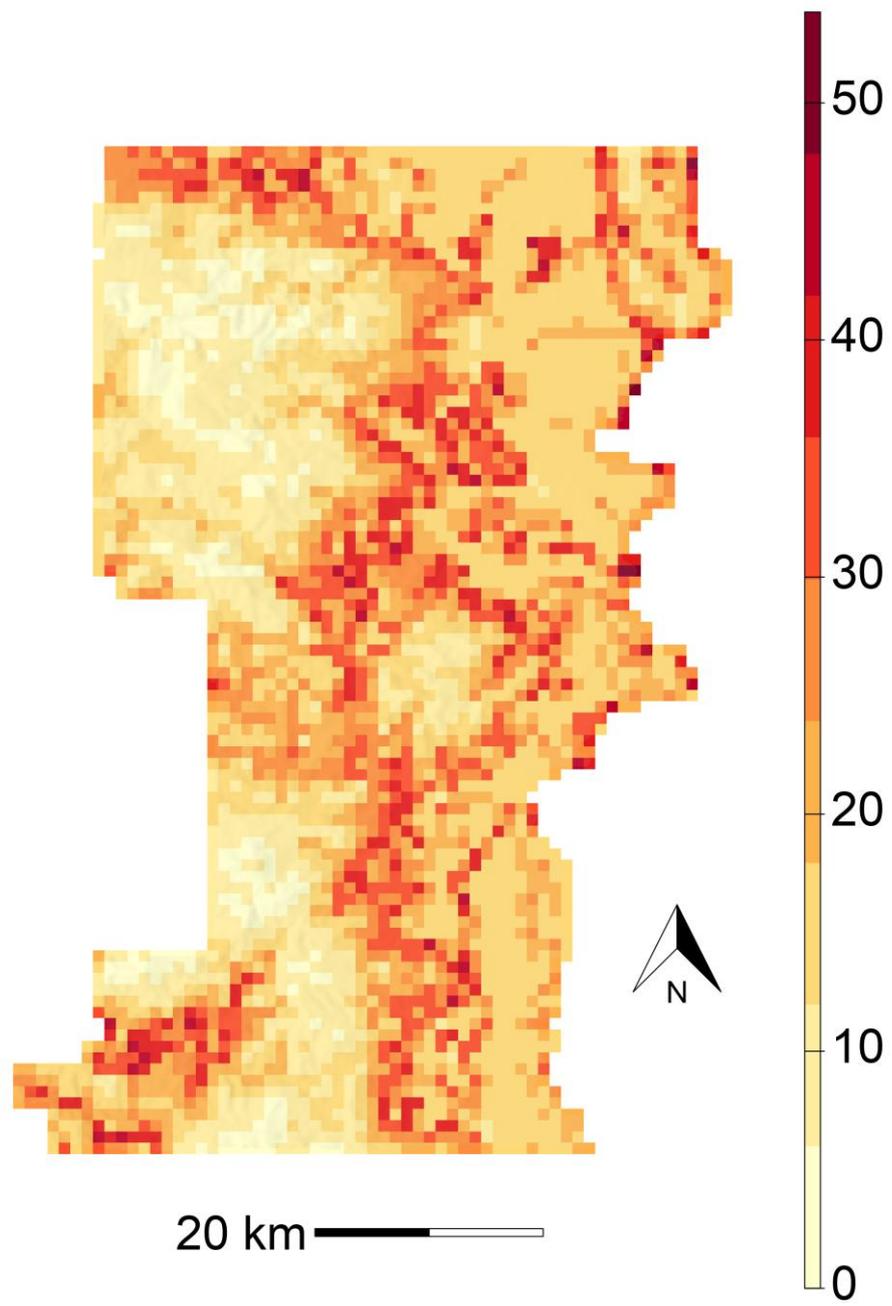

**Supporting Information.** Kéry, M., Royle, J.A., Hallman, T., Robinson, W.D., Strebel, N., Kellner, K.F., 2023. Integrated distance sampling models for simple point counts, *Ecology* xx.

Any use of trade, product, or firm names is for descriptive purposes only and does not imply endorsement by the U.S. Government.

# Appendix S1

This Appendix contains a **conceptual outline of the new class of IDS models** developed in the above-mentioned paper. In its three sections, we give an overview of the types of data that can be utilized in the integrated model (Section S1), present the statistical model that we propose for the joint analysis of these data in terms of the assumed processes that it contains (Section S2), and we close with some further comments and an outlook on possible further developments (Section S3).

**Section S1: Overview of the three types of data**
As currently described in the article and implemented in the function IDS() in the R package unmarked, three closed-population data types can be jointly analyzed: distance sampling (DS) data, an obligate ingredient of every IDS model, and then one or both of two further kinds of distance-free data: simple point count (PC) data or detection/nondetection (DND) data. All three are assumed to be collected by a stationary observer, i.e., during point-based surveys. However, the extension to data collected along line transects, i.e., by a moving observer, is conceptually straightforward.

Figure S1 represents such point-based survey data in a hypothetical landscape inhabited by 100 birds (red circles) preferring yellow over blue. Nine stationary observers (black triangles) survey birds out to some maximum distance (black circles) and collect either distance sampling data (left), distance-free point counts (middle), or distance-free detection/nondetection data (right). To avoid clutter in these maps, we here assume perfect detection, but in reality some individuals would typically be missed in distance sampling and point count surveys, and likewise for presences in detection/nondetection surveys. We also note that we depict all three survey types as being conducted from the identical nine locations. In practice, however, different locations (and also maximum survey distances) will typically be chosen, either by design or opportunistically (see also Figure S2).

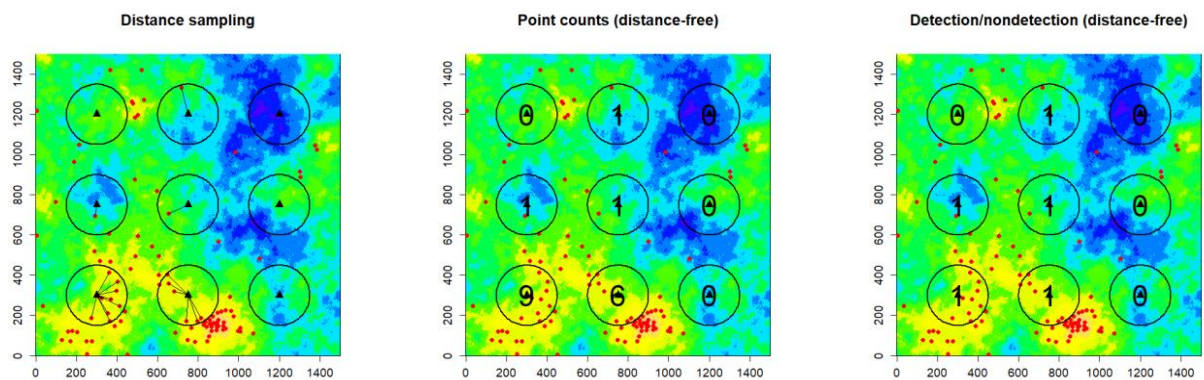

**Figure S1:** Schematic of three types of point-based surveys in a hypothetical landscape, showing how the same underlying process, combined with different survey methods, gives rise to the three data types that we combine in an IDS model.

## Section S2: Overview of IDS models

An IDS model is a probabilistic representation of the processes that we imagine underlie these three data types: (1) the *completely shared state process*, (2) a *partially shared observation process*, and (3) what can be called an *aggregation/summary process*; the latter is completely different for each data type.

### (1) Shared latent state process: a point process summarized by density per unit area

Figure S2 shows the shared state process, red points are again bird locations. Triangles are the locations of a human observer doing "point-transect" distance sampling surveys (left), distance-free point counts (middle), or distance-free detection/nondetection surveys (right). Survey locations are here shown to be different among survey types, as they will most often be in practice. Circles indicate the maximum distance out to which birds are surveyed, and this truncation distance is here depicted to be different among survey types, as will also typically be the case in the real world. The parameters linking density to the environmental covariates are assumed to be identical for all data types combined in an IDS analysis. Therefore, we call this a completely *shared* latent state process, both structurally (i.e., in terms of the density model) as well as in terms of the actual parameter values linking density with the landscape.

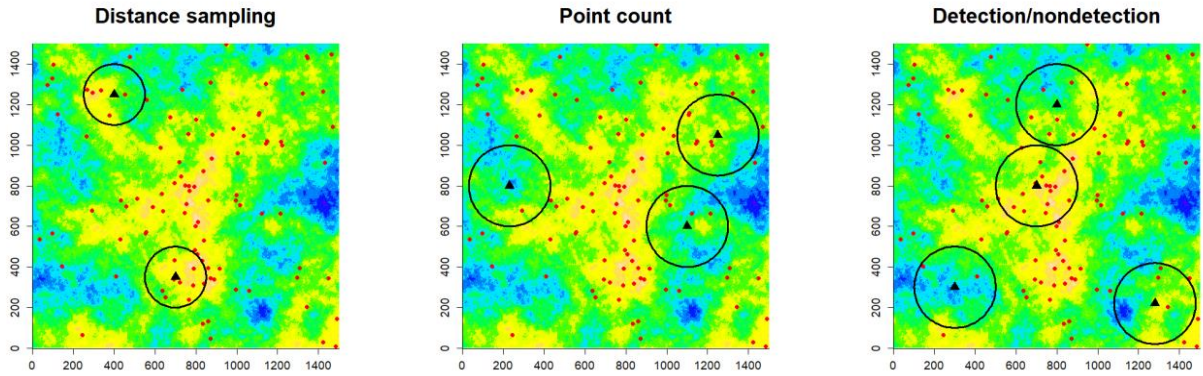

**Figure S2:** Another schematic of three types of point-based surveys in the same hypothetical landscape, this time emphasizing two typical features of real data sets fused in an IDS model: different survey locations and detection radii/truncation distances.

We can specify the state model in an IDS model as the outcome of a discrete-valued distribution such as a (possibly zero-inflated) Poisson or Negative binomial:

$$N_i \sim \text{Poisson}(\lambda_i)$$

### (2) Partially shared observation process: a distance sampling detection model

*For all data types, the fundamental observation process in an IDS model is a distance sampling detection function* (Figure S3). This is a novel and key assumption of our models and a distinguishing feature of our models from other types of integrated models that contain distance sampling as one of their ingredients, such as Solymos et al. (2013), Farr et al. (2021), and Schmidt et al. (2022). But although this structural part of the model is shared by all data types, different detection function parameters can be estimated for each data type, as we show in Simulation 1 of our paper. The truncation distance is shown by the vertical dotted line in Figure S3. For all three data types, the average per-individual detection probability in the area associated with the distance

sampling, point count, or detection/nondetection surveys is obtained by integration of the detection function from 0 out to the known or selected truncation distance.

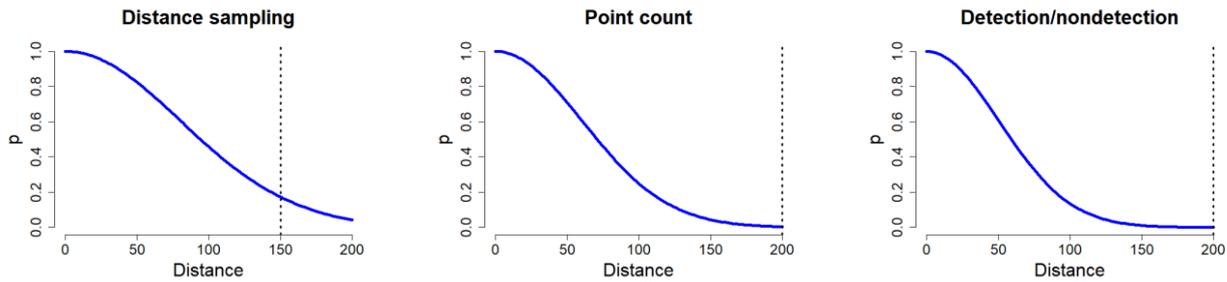

**Figure S3:** Schematic showing that key distinguishing feature of our integrated distance sampling models: an observation model represented by a detection function for each data type, and which may differ in terms of both its parameters and the truncation distance chosen when computing the detection probability.

**(3) Aggregation/summary process, rendering some information unavailable**
Between the distance sampling observation process and the actual data at hand, we can imagine an additional aggregation or summary process, which differs between the three data types. We emphasize that this third process generating the data used for an IDS model is somewhat hypothetical: we need this process to generate (i.e., simulate) data sets of the three kinds, but in practice, field observers are not actually first collecting distance sampling data and then discarding the distances to create simple counts. We invoke this additional process here since it clarifies the relationships among the three data types: it is required to explain how we get from the latent distance sampling observation model assumed for all three data types to the actual data for point count and detection/nondetection surveys.

| Distance sampling data | Point count data | Detection/nondetection data |
|---|---|---|
| • no aggregation or summary | • ignore/discard distance information<br>• counts of individuals, tally up the number of individuals detected<br>• alternatively, we could say that we simply collect no distance data and just count all individuals we detect | • ignore/discard distance information<br>• summarize resulting counts by a binary indicator function for counts >0<br>• alternatively, we could say that we collect neither distance nor count data, only detections and nondetections |

**(4) Input data for IDS model**
In the end, the shared state process (depicted in Figure S2), and the (partially shared) observation process (Figure S3), and the particular aggregation/summary processes combine to create three different types of data that form the input for an IDS model.

| Distance sampling | Point count | Detection/nondetection |
|---|---|---|

| Distance data for all individuals detected (e.g., 245 m, 128 m, 97 m) | Distance-free counts of unique individuals detected (e.g., 3, 0) | Distance-free detection/nondetection data (e.g., 1, 0) |

**(5) Observation models for observed data**
The typical observation models in an IDS, and those in the new `IDS()` function in R package `unmarked`, are a multinomial for the vector of binned distance counts **y** for the distance sampling data (Royle *et al.* 2004), a binomial (as in a binomial mixture model; Royle 2004) for point counts, and a Bernoulli distribution (as in the model of Royle & Nichols 2003) for detection/nondetection data.

| Distance sampling | Point count | Detection/nondetection |
|---|---|---|
| $\boldsymbol{y}_i \sim \text{Multinomial}(N_i, \boldsymbol{\pi})$ | $y_i \sim \text{Binomial}(N_i, \bar{p})$ | $y_i \sim \text{Bernoulli}(1 - (1 - \bar{p})^{N_i})$ |

In the multinomial observation model for the binned distance sampling data, the cell probabilities in $\boldsymbol{\pi}$ are given by an integration of the detection function over the chosen distance breaks that define the distance bins. In the binomial/Bernoulli observation model for the point counts and the detection/nondetection data, $\bar{p}$ is obtained as the integral of the detection function from 0 out to either a truncation distance (if there is one), or else to a sufficiently large distance that is chosen so that no individual can be detected further than that.

**Section S3: Further comments and outlook on future model extensions**

**(1) Shared state process**
In our IDS models the state process is assumed to be identical among all data sets combined in a single analysis. That means that the parameters relating density and the environmental covariates must be identical for all data sets. In practice this will mean that all surveys are conducted in the same region, as depicted in Figure S2. Sometimes, however, we may want to combine data types in an IDS model that do not show an adequate geographical interspersion, but rather may come from separate regions. Whether the integration of these data makes sense or not is a judgment call exactly as in any other study where we combine in a single analysis data from multiple places or from different times.

**(2) Observation process**
*The key idea of an IDS model is to treat distance sampling as the fundamental observation process for all three data types.* That is, we posit a distance-related, monotone decline of detection probability with increasing observer-bird distance and we describe it with a detection function such as the half-normal (though other detection functions could be chosen instead; Buckland et al. 2015; Chapter 8 in Kéry & Royle, 2016). The parameters of this detection function can be directly estimated from distance sampling data alone, but not from point count or detection/nondetection data alone. However, as part of an IDS model, i.e., in combination with distance sampling data and the assumption of a shared state process, the parameters of this ("latent") detection function become estimable also for point count and detection/nondetection data, as we demonstrate in Simulations 1–3 in our paper.

**(3) Availability process**
This conceptual outline describes how a basic IDS model "work". That is, an IDS model that is based conceptually on what is called conventional distance sampling (CDS; Buckland *et al.* 2015), where either availability probability is equal to 1 (e.g., every individual present in the surveyed area around each point is likely to sing at least once during a survey) or else where all data types combined in an IDS model stem from surveys with identical duration; in this case availability can be assumed to be identical for all. In many surveys of songbirds, the availability process will be governed mainly by the singing rate of an individual. Sometimes, we will have extra-information which enables us to estimate the parameters of this availability process in addition to those of the other processes described above; we show this in Simulation 5 of the paper. Such extra information may come in the form of varying survey duration, removal counts (Solymos et al. 2013), or external estimates of singing rates from outside of a study.

**(4) Temporal replicates (within "season" or "among seasons") and open populations**
In our paper we develop basic IDS models, with or without availability, for a static population. Incorporation of temporal replicates is the subject of future work (see also the Discussion of our paper). In principle, such extensions can be achieved by integrating the appropriate more complex likelihoods in the same manner as done in this paper. For example, Chandler et al. (2011) for repeated counts (or DS data) collected over time subject to temporary emigration, or the fully dynamic model of Dail and Madsen (2010). To the extent that these models have been described previously, there are no technical challenges to this generalized IDS model, although the code to implement them may be challenging even for specific cases let alone for general application.

**Literature cited**

Buckland, S.T., Rexstad, E.A., Marques, T.A., Oedekoven, C.S. 2015. *Distance sampling: methods and applications.* Springer, Cham, Switzerland.

Chandler, R.B., Royle, J.A., King, D.I. 2011. Inference about density and temporary emigration in unmarked populations. *Ecology*, 92, 1429–1435.

Dail, D., Madsen, L. 2011. Models for estimating abundance from repeated counts of an open population. *Biometrics*, 67, 577–587.

Farr, M.T., D.S. Green, K.E. Holecamp, E.F. Zipkin. 2021. Integrating distance sampling and presence-only to estimate species abundance. *Ecology*, 102, e03204.

Kéry, M., Royle, J.A. 2016. *Applied hierarchical modeling in ecology—Modeling distribution, abundance and species richness using R and BUGS. Volume 1: Prelude and Static Models.* Elsevier / Academic Press.

Royle, J.A. 2004. N-mixture models for estimating population size from spatially replicated counts. *Biometrics*, 60, 108–115.

Royle, J.A., Dawson, D.K., Bates, S. 2004. Modeling abundance effects in distance sampling. *Ecology*, 85, 1591–1597.

Royle, J.A., Nichols, J.D. 2003. Estimating abundance from repeated presence-absence data or point counts. *Ecology*, 84, 777–790.

Schmidt, J.H., Wilson, T.L., Thompson, W.L., Mangipane, B.A. 2022. Integrating distance sampling survey data with population indices to separate trends in abundance and temporary immigration. *Journal of Wildlife Management*, 86, e22185.

Sólymos, P., Matsuoka, S.M., Bayne, E.M., Lele, S.R., Fontaine, P., Cumming, S.G., Stralberg, D., Schmiegelow, F.K.A., Song, S.J. 2013. Calibrating indices of avian density from non-standardized survey data: making the most of a messy situation. *Methods in Ecology and Evolution*, 4, 1047–1058.

Any use of trade, product, or firm names is for descriptive purposes only and does not imply endorsement by the U.S. Government.

# Appendix S2

This Appendix contains the following supporting material for our paper.
- Section S1: An additional simulation study exploring the identifiability of our model under a wide range of parameter values.
- Section S2: Data selection criteria for the case study (AMRO in OR).
- Section S3: Frequentist operating characteristics of the IDS estimators in Simulations 1–4
- Section S4: Marginal posterior summaries of parameters in the IDS3 model fit to the American Robin data in the case study.

**Section S1:** *Simulation 1B: Frequentist operating characteristics of IDS estimators with DS + PC data under an even wider range of conditions*

To gauge the performance of the IDS1 model under a very wide range of conditions, we simulated and analysed data according to a response surface design, where we varied each of four settings independently: average density (0.1–5), detection function sigma for both DS and PC data (20–120m), and DS truncation distance (100–300m; we assumed unlimited distance for the PC data). We used JAGS to fit the IDS model to 1000 data sets comprising 250 DS and 1000 PC sites.

    Results suggested absence of bias, good coverage and thus excellent frequentist operating characteristics of the IDS1 model (DS + PC data) under the wide range of conditions simulated (Fig. S1, Table S2), thus confirming identifiability of the model. Relative bias was <<1% for the two sigma parameters and for the density, while for site-level abundances, it was about 2%. Coverage of Credible intervals (CRIs) varied between 0.90 and 0.94.

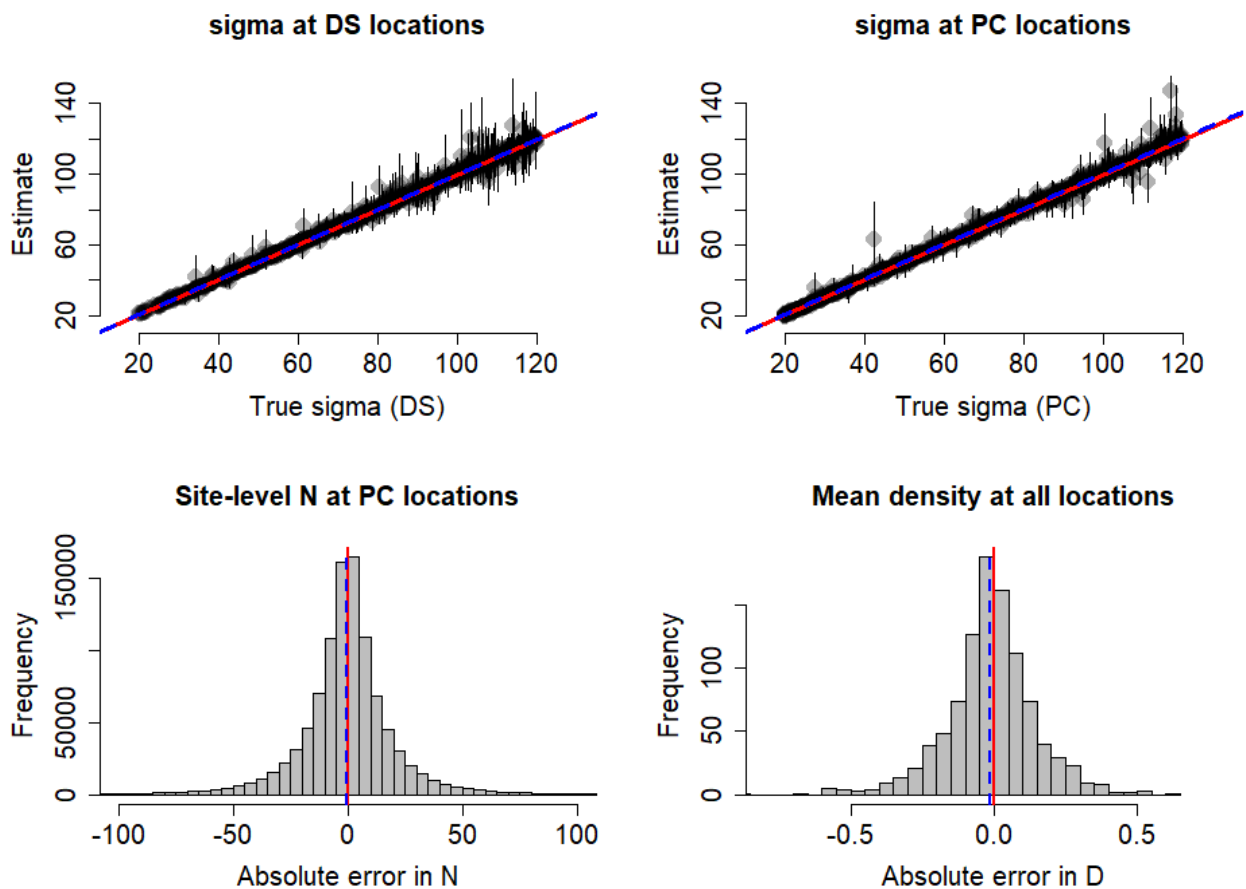

Figure S1: Sampling distributions of estimators in an IDS model with DS + PC data based on 1000 simulated data sets under a response-surface design (Simulation 2), see text for details. Top: posterior means and 95% CRIs, bottom: frequency distributions of posterior means. Red denotes truth or absence of estimation error, dashed blue shows mean of estimates. Compare also with Table S2.

**Section S2: eBird Selection Criteria for PC/DND data in Case study of paper**

**Source:** eBird Database

**Version:** ebd_US-OR_201101_201712_relNov-2017

**Years:** 2011 - 2017

**Dates:** Breeding season (April 30–July 11)

**Spatial Extent:** Benton and Polk Counties, Oregon

**Complete Checklists:** Yes (i.e., only complete checklists included)

**Presence-only Data:** Removed all checklists with just X

**Protocol:** Stationary

**Location Type:** Personal

**Time of Day:** Sunrise to 7 hours after sunrise

**Duration:** 3 - 30 minutes

**Grain of Geographic Sampling:** 200 m

**Checklists in Geographic Sampling:** 1 randomly selected per grid cell

**Section S3: Frequentist operating characteristics of the IDS estimators in Simulations 1–4**

**Table S1:** Simulation 1: Absolute bias, % relative bias, and 95% CRI coverage of the estimators (posterior means) of the half-normal scale parameter (sigma, $\sigma$) in the distance-sampling (DS) and the point count (PC) parts of the data, and of the site-level abundance (N) at the point-count sites under IDS models 1 and 2. An asterisk means that in this statistic all sites with true zero abundance were discarded due to division by zero. Sample size is 1000 sets of estimates from converged runs of the MCMC algorithm for each. Compare also with Figure 1 in main paper.

|                      | IDS1 (DS + PC) | | | IDS2 (DS + DND) | | |
|----------------------|-----------|--------|--------------|-----------|--------|--------------|
|                      | Abs. bias | %Bias  | CRI coverage | Abs. bias | %Bias  | CRI coverage |
| sigma (DS)           | 0.26      | 0.26   | 0.94         | 0.2132    | 0.21   | 0.95         |
| sigma (PC or DND)    | 0.13      | 0.09   | 0.94         | -0.0804   | -0.07  | 0.94         |
| Site N               | -0.11     | 1.10$^*$ | 0.94       | -0.1008   | 1.17   | 0.94         |

**Table S2:** Simulation 1B: Absolute bias, % relative bias, and 95% CRI coverage of the estimators (posterior means) of the half-normal scale parameter (sigma, $\sigma$) in the distance-sampling (DS) and the point count (PC) parts of the data, of the site-level abundance (*N*) at the point-count sites, and of mean density under the IDS model 1 (which combines DS + PC data). Asterisk: in this statistic all sites with zero true abundance were lost due to division by zero. Sample size is 1000 sets of estimates from converged runs of the MCMC algorithm. Compare also with Figure S1 above.

|               | Abs. bias | %Bias    | CRI coverage |
|---------------|-----------|----------|--------------|
| sigma (DS)    | 0.20      | 0.37     | 0.93         |
| sigma (PC)    | 0.23      | 0.29     | 0.94         |
| Site-level *N*| -0.80     | 2.14$^*$ | 0.94         |
| Density       | -0.01     | -0.45    | 0.90         |

**Table S3:** <u>Simulation 2:</u> Absolute bias, % relative bias, and 95% CRI coverage of the estimators (MLEs) in IDS1 combining distance-sampling (DS) and point count (PC) data and where the covariate effects in the detection functions of the DS and the PC data differ (Simulation 2a). Left: independent covariates are simulated in the density and the detection function parts of the model; right: a single covariate affects density as well as the two detection functions (Simulation 2b). Sample size is 977 (left) and 934 (right) valid model fits; see main text for more explanation. Compare also with Figure 2 in main paper.

|  | Simulation 2a: Independent density and detection covariates | | | Simulation 2b: A single covariate affects both density and detection | | |
|---|---|---|---|---|---|---|
|  | Abs. bias | %Bias | CI coverage | Abs. bias | %Bias | CI coverage |
| Density (intercept) | -0.001 | -0.09 | 0.95 | 0.002 | 0.16 | 0.95 |
| Density (slope) | 0.00 | 0.00 | 0.94 | 0.00 | 0.07 | 0.96 |
| sigma DS (intercept) | 0.18 | 0.18 | 0.95 | -0.05 | -0.05 | 0.95 |
| sigma DS (slope) | 0.00 | 0.00 | 0.94 | 0.00 | 0.00 | 0.94 |
| sigma PC (intercept) | 0.12 | 0.08 | 0.95 | 0.06 | 0.04 | 0.95 |
| sigma PC (slope) | 0.00 | 0.00 | 0.94 | -0.001 | 0.00 | 0.96 |

**Table S4:** Simulation 3: Absolute bias, % relative bias, and 95% CRI coverage of the estimators (MLEs) of the shared intercept and the slope of density ($\lambda$), and of the scale parameter (sigma, $\sigma$) in the distance-sampling (DS) and the point count (PC) parts of the data under the IDS model 1 (which combines DS + PC data), when between 1 and 100 DS sites are added in the analysis of 200 PC sites. Original sample size is 1000 data sets at each added value of DS sites. Compare also with Figure 3 in main paper.

|  | 1 | 20 | 40 | 60 | 80 | 100 |
|---|---|---|---|---|---|---|
| Valid cases | 515 | 915 | 962 | 981 | 989 | 998 |
| **$\lambda$ intercept** | | | | | | |
| Abs. bias | 0.533 | 0.002 | 0.005 | 0.001 | 0.002 | 0.000 |
| %Bias | 53.34 | 0.22 | 0.46 | 0.11 | 0.19 | 0.03 |
| CI coverage | 0.96 | 0.96 | 0.95 | 0.95 | 0.93 | 0.94 |
| **$\lambda$ slope** | | | | | | |
| Abs. bias | 0.001 | 0.002 | -0.001 | -0.000 | -0.000 | -0.000 |
| %Bias | 0.12 | 0.16 | -0.07 | -0.02 | -0.03 | -0.05 |
| CI coverage | 0.94 | 0.95 | 0.95 | 0.95 | 0.94 | 0.95 |
| **$\sigma^{DS}$** | | | | | | |
| Abs. bias | -5.714 | 0.743 | 0.224 | 0.310 | 0.069 | 0.209 |
| %Bias | -5.71 | 0.74 | 0.22 | 0.31 | 0.07 | 0.21 |
| CI coverage | 0.89 | 0.95 | 0.97 | 0.95 | 0.95 | 0.95 |
| **$\sigma^{PC}$** | | | | | | |
| Abs. bias | -2.487 | 0.320 | 0.022 | 0.127 | 0.037 | 0.072 |
| %Bias | -3.55 | 0.46 | 0.03 | 0.18 | 0.05 | 0.10 |
| CI coverage | 0.96 | 0.96 | 0.95 | 0.95 | 0.95 | 0.95 |

**Table S5:** Simulation 4: Absolute bias, % relative bias, and 95% CRI coverage of the estimators (MLEs) of an IDS model with availability: abundance intercept ($\lambda$), availability intercept (represented by singing rate $\phi$) and detection function scale parameter ($\sigma$) in the distance-sampling (DS) and the point count (PC) portions of the data. Compare also with Figure 4 in main paper. Results shown from those data sets that did not result in numerical failures when fitting the model using MLE with function IDS().

|  | Abs. bias | %Bias | CI coverage |
|---|---|---|---|
| **3000 DS + 1000 PC (nvalid = 930)** | | | |
| $\lambda$ (intercept) | 0.0141 | 14.09 % | 0.96 |
| $\phi$ (intercept) | 0.0657 | 9.82 % | 0.96 |
| $\sigma$ (DS) | 0.0040 | 0.004% | 0.93 |
| $\sigma$ (PC) | 0.5324 | 0.76 % | 0.96 |
| **3000 DS + 3000 PC (nvalid = 848)** | | | |
| $\lambda$ (intercept) | 0.0457 | 4.57 % | 0.95 |
| $\phi$ (intercept) | 0.0285 | 4.07 % | 0.95 |
| $\sigma$ (DS) | 0.0070 | 0.01% | 0.95 |
| $\sigma$ (PC) | 0.1538 | 0.22 % | 0.95 |
| **3000 DS + 6000 PC (nvalid = 997)** | | | |
| $\lambda$ (intercept) | 0.0282 | 2.82 % | 0.96 |
| $\phi$ (intercept) | 0.0206 | 2.37 % | 0.96 |
| $\sigma$ (DS) | -0.0711 | 0.07% | 0.94 |
| $\sigma$ (PC) | 0.0761 | 0.11 % | 0.96 |

**Section S4:**

**Table S6: Marginal posterior summaries of parameters in the IDS3 model fit to the American Robin data in the case study.**

Estimates based on a run of JAGS with 4 chains of 120000 iterations, adaptation = 10000 iterations, burn-in = 60000 iterations, and thin rate = 60, yielding 4000 total samples from the joint posterior. Columns give the posterior means and standard deviations, and the 2.5, 50 and 97.5% quantiles.

```
                       mean         sd      2.5%        50%      97.5%
Model for the half-normal detection function (parameter sigma)
-------------------------------------------------------------
mean.sigma[1]         0.049     0.0031     0.043      0.049      0.055
mean.sigma[2]         0.047     0.0041     0.040      0.047      0.056
alpha0[1]            -3.020     0.0634    -3.147     -3.021     -2.894
alpha0[2]            -3.053     0.0874    -3.224     -3.053     -2.877
alpha1               -0.766     0.0570    -0.878     -0.766     -0.651
alpha2               -0.384     0.1173    -0.611     -0.384     -0.151
sd.eps[1]             0.204     0.0169     0.171      0.204      0.236
sd.eps[2]             0.471     0.0453     0.384      0.470      0.561

Model for the availability parameter phi
----------------------------------------
mean.phi              0.271     0.0839     0.145      0.263      0.458
gamma0               -1.348     0.2856    -1.934     -1.334     -0.781
gamma1                0.400     0.4573    -0.057      0.218      1.609
gamma2                0.571     0.6088    -0.053      0.387      1.958
gamma3               -0.251     0.1266    -0.533     -0.232     -0.067
gamma4                0.173     0.1364     0.013      0.138      0.488

Model for density (parameter lambda)
------------------------------------
mean.lambda          32.063     7.9117    20.742     30.386     51.590
beta0                 3.440     0.2339     3.032      3.414      3.943
beta1                -1.074     0.2755    -1.609     -1.076     -0.538
beta2                -7.510     0.5232    -8.538     -7.508     -6.495
beta3                -0.676     0.0694    -0.810     -0.674     -0.541
beta4                 0.088     0.0214     0.046      0.089      0.129
sd.beta0              0.337     0.1469     0.157      0.303      0.699
ann.beta0[1]          3.152     0.1836     2.883      3.111      3.556
ann.beta0[2]          3.641     0.1630     3.423      3.591      4.022
ann.beta0[3]          3.708     0.1624     3.480      3.660      4.099
ann.beta0[4]          3.629     0.2477     3.231      3.588      4.157
ann.beta0[5]          3.132     0.2284     2.726      3.115      3.611
ann.beta0[6]          3.306     0.2547     2.850      3.284      3.841
ann.beta0[7]          3.381     0.2472     2.982      3.347      3.910

Posterior predictive checks of Goodness-of-Fit
----------------------------------------------
fitDS               901.726    23.4604   858.935    900.476    949.770
fitDS.new           956.579    26.1802   908.365    954.970   1010.771
fitPC               164.055    10.6127   144.084    163.906    185.450
fitPC.new           158.083     8.9977   140.582    158.125    175.799
fitDND              228.000     0.0000   228.000    228.000    228.000
fitDND.new          220.278    11.7641   197.000    220.000    243.000
bpvDS                 0.991     0.0945     1.000      1.000      1.000
bpvPC                 0.285     0.4517     0.000      0.000      1.000
```